\documentclass[aps,prb,twocolumn,superscriptaddress,floats,showpacs]{revtex4}
\usepackage{amssymb}
\usepackage{multirow}
\usepackage{graphicx}

\begin{document}

\title{The weak electronic correlations and absence of heavy Fermion state in KNi$_2$Se$_2$ }

\author{Q. Fan}
\author{X. P. Shen}
\affiliation{State Key Laboratory of Surface Physics, Department of
Physics,  and Advanced Materials Laboratory, Fudan University,
Shanghai 200433, P. R. China}
\affiliation{Collaborative Innovation Center of Advanced Microstructures, Nanjing 210093, P. R. China}
\author{M. Y. Li}
\author{D. W. Shen}\email{dwshen@mail.sim.ac.cn}
\author{W. Li}
\author{X. M. Xie}
\affiliation{State Key Laboratory of Functional Materials for Informatics,
Shanghai Institute of Microsystem and Information Technology (SIMIT),
Chinese Academy of Sciences, Shanghai 200050, P. R. China}

\author{Q. Q. Ge}

\author{Z. R. Ye}

\author{S. Y. Tan}

\author{X. H. Niu}

\author{B. P. Xie}

\author{D. L. Feng}\email{dlfeng@fudan.edu.cn}

\affiliation{State Key Laboratory of Surface Physics, Department of
Physics,  and Advanced Materials Laboratory, Fudan University,
Shanghai 200433, P. R. China}
\affiliation{Collaborative Innovation Center of Advanced Microstructures, Nanjing 210093, P. R. China}

\date{\today}

\begin{abstract}

We have studied the low-lying electronic structure of a new ThCr$_2$Si$_2$-type superconductor KNi$_2$Se$_2$ with angle-resolved photoemission spectroscopy. Three bands intersect the Fermi level, forming complicated Fermi surface topology, which is sharply different from its isostructural superconductor K$_x$Fe$_{2-y}$Se$_2$. The Fermi surface shows weak variation along the $k_z$ direction, indicating its quasi-two-dimensional nature. Further comparison with the density functional theory calculations demonstrates that there exist relatively weak correlations and substantial hybridization of the Ni 3$d$ and the Se 4$p$ orbitals in the low-lying electronic structure. Our results indicate that the large density of states at the Fermi energy leads to the reported mass enhancement based on the specific heat measurements. Moreover,  no anomaly is observed in the spectra  when entering the  fluctuating charge density wave  state reported earlier.

\end{abstract}


\maketitle

\section{introduction}
The iron-chalcogenide superconductors A$_x$Fe$_{2-y}$Se$_2$ (A=K, Rb, etc.) with superconducting transition temperature ($T_c$) up to 33~K have aroused a great deal of research interests \cite{KFeSe discovery, KFeSe AFM insulator, TlFeSe, KFeSe high TN, KFeSe phase diagram, Neutron, KFeSe review}. Compared with the iron pnictide superconductors, A$_x$Fe$_{2-y}$Se$_2$ exhibits some unique properties, such as the antiferromagnetically ordered insulator parent compound with Fe vacancy order \cite{KFeSe AFM insulator, KFeSe high TN, KFeSe phase diagram, Neutron}. Particularly, angle-resolved photoemission spectroscopy (ARPES) revealed that the Fermi surface of the superconducting phase is consisted of electron pockets only, raising a serious challenge to the prevalent Fermi surface nesting mechanism for the superconductivity in iron-based superconductors \cite{KFeSe ARPES1, KFeSe ARPES2}.

Recently, another chalcogenide superconductor KNi$_2$Se$_2$ was reported\cite{KNiSe polycrystal}, which shares the same crystal structure and almost the same lattice constants with K$_x$Fe$_{2-y}$Se$_2$, except that there are no vacancies in  KNi$_2$Se$_2$ \cite{KNiSe single crystal}. However, previous experiments indicated that KNi$_2$Se$_2$ displays a rich but sharply different phase diagram from that of K$_x$Fe$_{2-y}$Se$_2$. Based on the specific heat measurements and neutron pair-distribution-function analysis, KNi$_2$Se$_2$ was suggested to show a local charge density wave (CDW) fluctuating state above 20~K, which then enters a coherent heavy-Fermion state at lower temperatures. Eventually, it becomes superconducting below $\sim$0.8~K \cite{KNiSe polycrystal}. The large linear specific heat coefficient (or Sommerfeld coefficient, $\gamma$) was proposed to be due to the strong electron correlation and heavy Fermion behavior in KNi$_2$Se$_2$. More specifically, mix valency of the Ni atoms was proposed to induce the heavy effective band mass $m^*$. On the other hand, we note that the  band renormalization factor of BaNi$_2$As$_2$ (a superconductor with $T_c\sim$0.7~K) is merely $\sim$1.66, in the weak-interaction regime, and thus the conventional electron-phonon interaction mechanism was suggested to be the cause of the superconductivity therein \cite{BaNi2As2}. In an ionic picture, the Ni orbital configuration is 3$d^8$ for BaNi$_2$As$_2$, and 3$d^{8.5}$ for KNi$_2$Se$_2$. Consequently, the Hund's rule coupling and electron correlations should be weakened with the increasing number of $3d$ electrons in the latter compound \cite{ZirongPRX}, which is inconsistent with the proposed heavy Fermion scenario in KNi$_2$Se$_2$. To resolve this controversy and understand the unique properties of KNi$_2$Se$_2$, it is critical to study its electronic structure.

In this article, we have systematically studied the electronic structure of the single-crystalline KNi$_2$Se$_2$ with ARPES. There are two electron pockets around the $M$ point, a hole-like square pocket  surrounding the $\mathit\Gamma$ point, and a hole-like narrow race-track  pocket surrounding the $X$ point, respectively. The electronic structure of KNi$_2$Se$_2$ shows relatively weak $k_z$ dependence, indicative of its two-dimensional (2D) character. Further comparison between the  density functional theory (DFT) band calculations and ARPES data gives a  renormalization factor of about 1.54, which indicates that the electron correlations are weak. The coherent bands are present at high temperatures, and there is no incoherent-weight-to-coherent-band transition at low temperatures as in heavy Fermion materials. Instead, we find that  the Fermi velocities of several bands are relatively small  and thus  enhance the density of states (DOS) at the Fermi energy ($E_F$).  Our quantitative analysis shows that the large $\gamma$  of KNi$_{2}$Se$_{2}$ together with those of  BaNi$_{2}$As$_{2}$ and KFe$_{2}$As$_{2}$ (which has even larger $\gamma$), can be well accounted by the multiple large Fermi surfaces and relatively small Fermi velocities in these systems without invoking any heavy Fermion physics.

\section{experimental}
\begin{figure}[t]
\includegraphics[width=8.5cm]{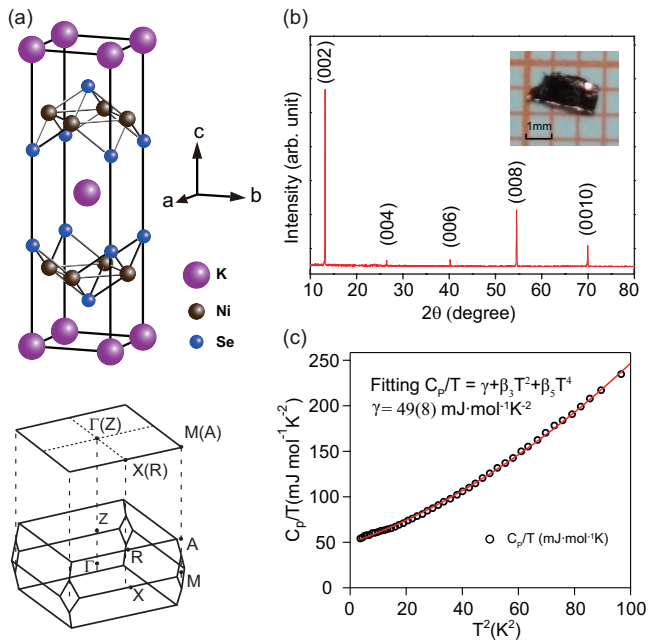}
\caption{(Color online) (a) Representative unit cell of KNi$_2$Se$_2$ (upper panel) and its Brillouin zone (lower panel). Note one tetragonal unit cell of KNi$_2$Se$_2$ contains two chemical formula units. For the convenience, the 2D  Brillouin zone is referred hereafter, which is the projection of the three-dimensional Brillouin zone. (b) X-ray diffraction patterns of KNi$_2$Se$_2$ single crystal after cleavage. The inset shows the photo of a typical single crystal. (c) The low-temperature specific heat data of KNi$_2$Se$_2$ divided by temperature versus temperature squared, revealing a large linear electronic contribution to the low-temperature heat capacity.}
\label{Characterization}
\end{figure}

High quality single crystals of KNi$_2$Se$_2$ were grown by self-flux method with nominal composition K:Ni:Se = 1:2:2. The mixture was loaded into the alumina crucible and then sealed in an argon-filled iron crucible. The entire assembly was kept at 1273 K  for 3 hours, and then cooled down to 873 K slowly at a rate of $\sim$4 K/h before shutting off the power. The samples are crystallized in the tetragonal ThCr$_2$Si$_2$-type structure with the space group $I4/mmm$, as shown in Fig.~\ref{Characterization}(a)\cite{KNiSe crystal structure}. The as-grown single crystals with a typical dimension of 2.5$\times$1.0$\times$0.1~mm$^3$ show flat shiny surface of pink color after cleavage [the inset of Fig.~\ref{Characterization}(b)]. The electron probe micro analysis (EPMA) measurements across samples with more than 10 points indicate that the composition is quite homogeneous, and the averaged stoichiometric ratio is determined to be KNi$_{2.06}$Se$_{2.01}$.  Since this determined composition is close to the stoichiometry, we will still designate the samples as KNi$_2$Se$_{2}$ for convenience hereafter. In Fig.~\ref{Characterization}(b), X-ray diffraction measurements show that only the series of ($00l$) narrow reflection peaks appear, suggestive of the good crystalline quality. The low-temperature specific heat [Fig.~\ref{Characterization}(c)] can be well fitted by the formula $C_{p}/T=\gamma+\beta_{3}T^{2}+\beta_{5}T^{4}$ in the temperature range from 1.8~K to 10~K, and the resulting large Sommerfield coefficient [$\gamma$=49.8 mJ mol$^{-1}$ K$^{-2}$] is in good agreement with previous reports \cite{KNiSe polycrystal,KNiSe single crystal}. For comparison, we note that the  corresponding $\gamma$  is about 13.2 mJ mol$^{-1}$ K$^{-2}$  for BaNi$_2$As$_2$ \cite{BaNiAs HC}, and about 94.3 mJ mol$^{-1}$ K$^{-2}$ for KFe$_2$As$_2$ \cite{KFeAs HC}.

ARPES measurements were performed at (1) Beamline 7U of the UVSOR synchrotron facility with a MBS A-1 electron analyzer, (2) Beamline 28A of
Photon Factory (PF), KEK, Tsukuba, with a Scienta SES-2002 analyzer, and (3) the in-house system equipped with an SPECS UVLS helium discharging lamp and VG Scienta R4000 electron analyzer. The overall energy resolution was set to 15-30~meV depending on the photon energy, and the typical angular resolution was 0.3$^\circ$. Samples were cleaved \emph{in situ} and then measured under ultrahigh vacuum better than 6$\times$10$^{-11}$ mbar.  The sample surfaces were stable and did not show any sign of degradation during the measurements.

The first-principles calculations were implemented in the VASP code \cite{VASP}. The plane wave basis method and the Perdew-Burke-Ernzerhof \cite{PBE} exchange correlation potential have been used. Throughout the calculations, a $500$ eV cutoff in the plane wave expansion and a $12\times 12\times 12$ Monkhorst-Pack $\vec{k}$ grid were chosen to ensure the calculation with an accuracy of $10^{-5}$ eV. In our calculations, the crystal structure and lattice constants were taken from the experimental values\cite{KNiSe polycrystal,KNiSe crystal structure}.

\section{Experimental results}

\begin{figure*}
\begin{minipage}{0.7\textwidth}
\includegraphics[width=11.5cm]{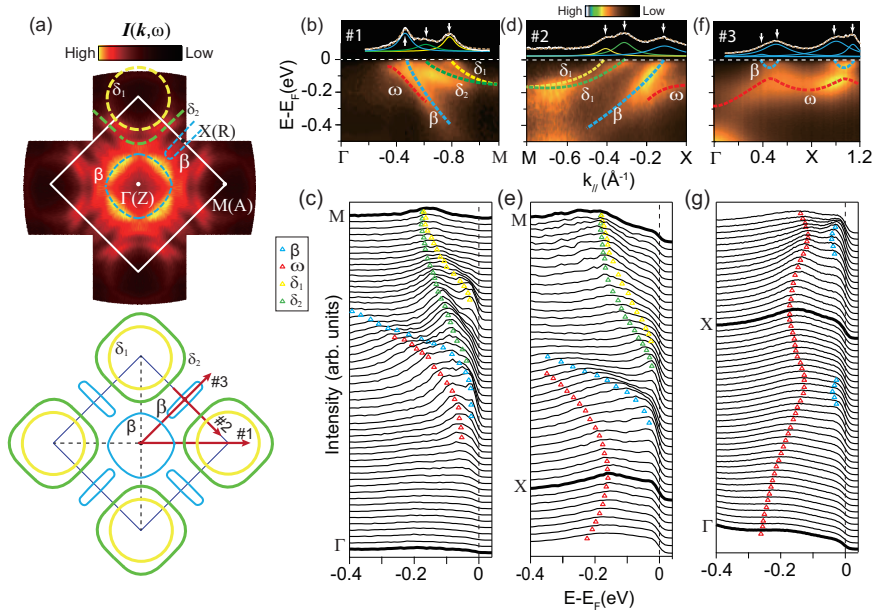}
\end{minipage}%
\begin{minipage}{0.3\textwidth}
\caption{(Color online) Electronic structure of KNi$_{2}$Se$_{2}$ measured at 15~K. (a) Photoemission intensity map integrated over [$E_F$ -10~meV, $E_F$+10~meV] (upper panel) and the 2D projected Brillouin zone of KNi$_{2}$Se$_{2}$ (lower panel). It was obtained through mirroring the data with respect to both $k_x$ and $k_y$ axes. (b) The photoemission intensity along cut \#1 ($\mathit\Gamma$~-~$M$  direction). (c) Selected EDCs for data in (b). (d) and (e), (f) and (g) show the similar data as in (b) and (c), but along cut \#2 and cut \#3, respectively. The data in panel (a) were taken with randomly polarized 21.2~eV photons from a helium discharge lamp, and the other data were taken with circularly polarized 75~eV photons at KEK.}
\label{FS}
\end{minipage}
\end{figure*}

The photoemission intensity map of KNi$_{2}$Se$_{2}$ at 15~K is shown in Fig.~\ref{FS}(a), which is overlaid on the projected 2D Brillouin zone.  The resulting Fermi surface (FS) consists of one square-like Fermi pocket around the zone center, two Fermi pockets around zone corner, and one narrow race-track pocket extending to the middle of zone boundaries, as highlighted by the colored dashed lines. This complicated FS is a direct evidence of the multi-band behavior in this compound, consistent with previous Hall effect measurement \cite{KNiSe single crystal} and theoretical calculations \cite{KNiSe band calculation1, KNiSe band calculation2}. Although this multi-band character is similar to iron-based superconductors, the detailed FS topology of KNi$_{2}$Se$_{2}$ is distinct from that of K$_x$Fe$_{2-y}$Se$_2$, which only exhibits circular electron-like pockets around the zone corner\cite{KFeSe ARPES1, KFeSe ARPES2}.

\begin{figure*}[t]
\includegraphics[width=12cm]{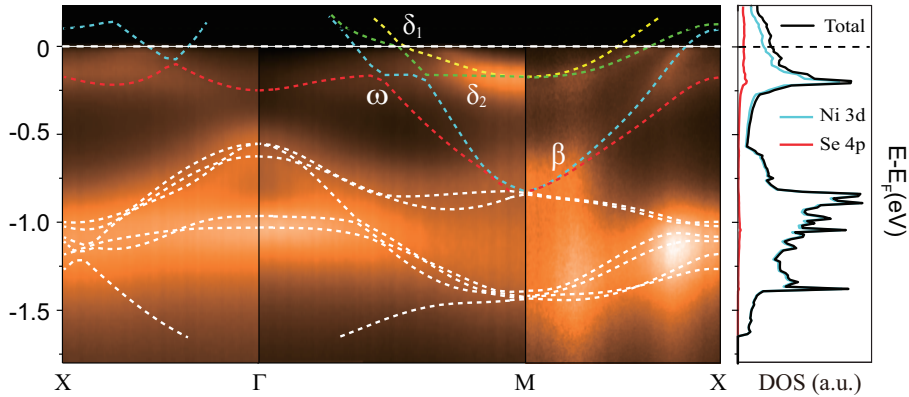}
\caption{(Color online) (Left) Comparison of the ARPES experimental  band
structure and the DFT calculation results along the high symmetry directions in KNi$_{2}$Se$_{2}$. (Right) Calculated density of states of KNi$_{2}$Se$_{2}$. Blue and red curves correspond to Ni and Se partial DOS, respectively.}
\label{DFT}
\end{figure*}

Now we examine the band dispersions of  KNi$_{2}$Se$_{2}$ along several high-symmetry directions as indicated in the lower panel of Fig.~\ref{FS}(a). We chose 75~eV photons to probe the band structure near the $\Gamma$XM plane of KNi$_{2}$Se$_{2}$  considering an inner potential of 15~eV as discussed below.
Along $\mathit\Gamma$~-~$M$ [cut \#1 in Fig.~\ref{FS}(b)], three Fermi crossings can be identified by fitting the corresponding momentum distribution curve (MDC) at $E_F$. Moreover, the corresponding band dispersions can be tracked with the peaks in the energy distribution curves (EDCs) [Fig.~\ref{FS}(c)], and they are assigned as $\beta$, $\delta_1$, and $\delta_2$, respectively. Among them, the hole-like $\beta$ band encloses the $\mathit\Gamma$ point, forming the square-shaped Fermi pocket; while, the $\delta_1$ and $\delta_2$ bands seem to be degenerate at the $M$ point with the band bottoms at 180~meV, developing two electron pockets around the zone corner. Besides, there is another weak feature ($\omega$) around the $\mathit\Gamma$ point, whose band top is about 50~meV below  $E_F$. Along $M$~-~$X$ [ cut \#2 in Fig.~\ref{FS}(d)], three bands cross $E_F$ as well, which are further confirmed by the peaks in the corresponding MDC fitting and EDCs [Fig.~\ref{FS}(e)]. Taking into account the bottom positions of these bands, we can infer that these two electron-like bands around $M$ are  the $\delta_1$ and $\delta_2$ bands, respectively. Based on the band calculations presented in Fig.~\ref{DFT} later, the band near $X$ (marked by blue dashed lines) is $\beta$.

Along $\mathit\Gamma$~-~$X$, data in Figs.~\ref{FS}(f) and \ref{FS}(g) show a shallow electron-like band (blue dashed line). Based on the analysis presented in Fig.~\ref{DFT} later, this feature is actually  contributed mostly by the $\beta$ band. Such a shallow electron-like band is due to a pronounced  hybridization between $\beta$ and $\omega$ shown here. However, it does not develop into a closed electron pocket. Instead, it forms the narrow race-track hole-like Fermi pocket ($\beta$) surrounding $X$.

\begin{figure}[t]
\includegraphics[width=8.5cm]{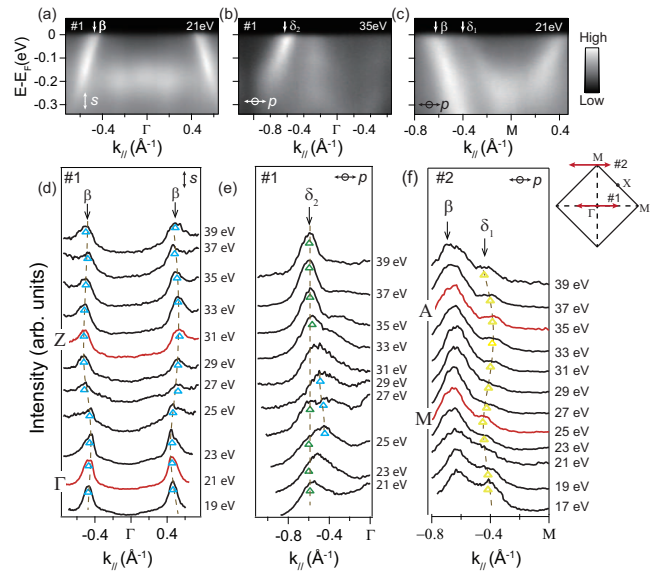}
\caption{(Color online) (a) The photoemission intensity taken along cut \#1  with $s$ polarized 21~eV photons. (b) The photoemission intensity  taken along cut \#1 with $p$ polarized 35~eV photons. (c) is the same as (a) but along cut \#2. (d) and (e) are MDCs at $E_F$ taken with different photon energies along cut \#1 in $s$ and $p$ polarizations, respectively. (f) is the same as (d) but taken along cut  \#2 in $p$ polarization. Data were taken at Beamline 7U of UVSOR, and the temperature was 66~K.}
\label{Kz}
\end{figure}

To fully reveal the Fermi surface topology in the three-dimensional Brillouin zone, we have performed detailed photon energy dependent ARPES measurements. Along cut \#1, only $\beta$ band shows up in the $s$ polarization [Fig.~\ref{Kz}(a)]. The Fermi momenta ($k_F$'s) of $\beta$ are determined by peak positions in MDCs in [Fig.~\ref{Kz}(d)]. Taking the inner potential of 15~eV to estimate the $k_z$'s for different photon energies, we notice that $k_F$'s of $\beta$ move periodically from $\mathit\Gamma$ (probed with 21~eV photons) to $Z$ (probed with 31~eV photons). Our data could cover more than half of the Brillouin zone along the $k_z$ direction. On the other hand, $\delta_2$ is clearly resolved in the $p$ polarization [Fig.~\ref{Kz}(b)], which shows negligible $k_z$ dispersion as shown in Fig.~\ref{Kz}(e).  Along cut \#2, $\beta$ and $\delta_1$ band show up in the $p$ polarization [Fig.~\ref{Kz}(c)].  As shown in Fig.~\ref{Kz}(f),  there is some  dispersion along $k_z$ for the $\delta_1$ band. In general, the $k_z$ dispersions of these bands are not strong, manifesting the 2D nature of  KNi$_{2}$Se$_{2}$. Our result is consistent with the previous theoretical calculations except for the hole-like pocket centered at the $\mathit\Gamma$ point, which was suggested to be more three-dimensional-like\cite{KNiSe band calculation1, KNiSe band calculation2}. This discrepancy might be partially attributed to the poor $k_z$ resolution in ARPES experiments.

\section{Discussions and Conclusions}

The measured quasi-2D electronic structure of KNi$_{2}$Se$_{2}$  is compared with the band structure obtained  from our first-principles calculations. In Fig.~\ref{DFT}, the  calculated band  (dashed lines) are scaled and appended onto the photoemission intensity plots along the three high symmetry directions. Here, a renormalization factor of about 1.54 to the calculations leads to a remarkable agreement with the experimental data. Almost all characteristic dispersions, even the small electron-like feature along $\mathit\Gamma$~-~$X$ direction, can be reproduced well by the calculations. This relatively small renormalization factor demonstrates that the electron correlations in this system are rather weak. Furthermore, we present the calculated total DOS and the projected DOS of KNi$_{2}$Se$_{2}$   in the right panel of Fig.~\ref{DFT}. Near $E_F$, there are some contribution from the Se 4$p$ orbitals   in addition to the dominating contribution from Ni 3$d$ orbitals.

The small renormalization factor found here for  KNi$_{2}$Se$_{2}$ is similar to that of  BaNi$_{2}$As$_{2}$ ($\sim$1.66) \cite{BaNi2As2}, which suggests that electron correlations are weak in both systems. This is consistent with  their Ni orbital configurations, \textit{i.e.} 3$d^8$ for BaNi$_2$As$_2$, and 3$d^{8.5}$ for KNi$_2$Se$_2$. Hund's rule coupling is the main source of correlations in these materials, and
it has been shown recently that the correlations are reduced dramatically from 3$d^6$ to 3$d^7$ with electron doping, and superconductivity  diminishes with weakened correlations \cite{ZirongPRX}. Therefore, compared with the iron-based superconductors, the even higher 3$d$ occupation and weaker correlations in these Ni-based compound suggest that the  superconductivity is most likely due to the electron-phonon coupling, similar to that in the multi-band superconductor MgB$_2$ \cite{1MgB2, 2MgB2}.

\begin{figure*}[t]
\includegraphics[width=16cm]{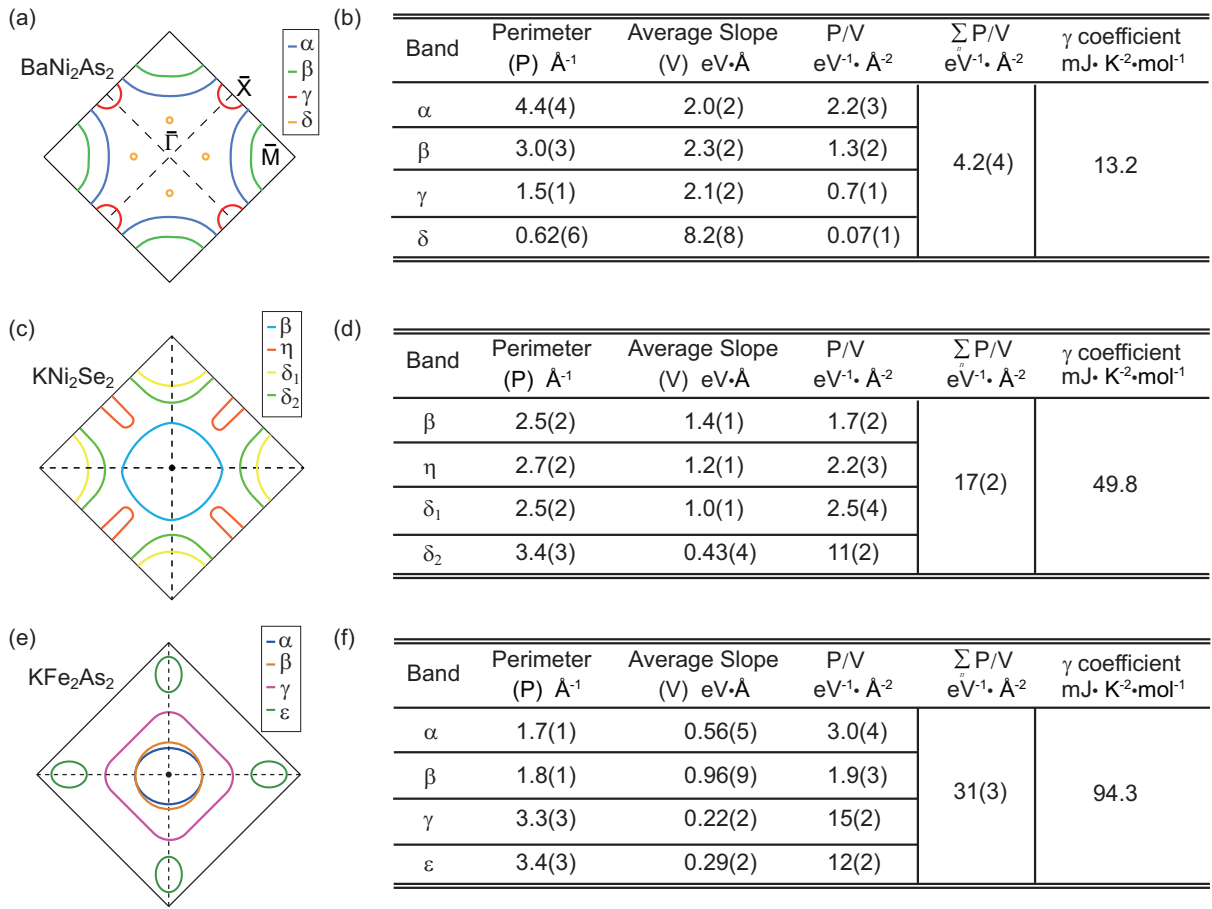}
\caption{(Color online) (a) Sketch of Fermi surface for BaNi$_{2}$As$_{2}$, reproduced from Ref. \cite{BaNi2As2}. (b) The DOS estimation of each band in BaNi$_{2}$As$_{2}$ system. (c) and (d), (e) and (f) are the same as in panels (a) and (b), but for KNi$_{2}$Se$_{2}$ and KFe$_2$As$_{2}$, respectively. }
\label{DOSestimation}
\end{figure*}

\begin{figure}[t]
\includegraphics[width=8.5cm]{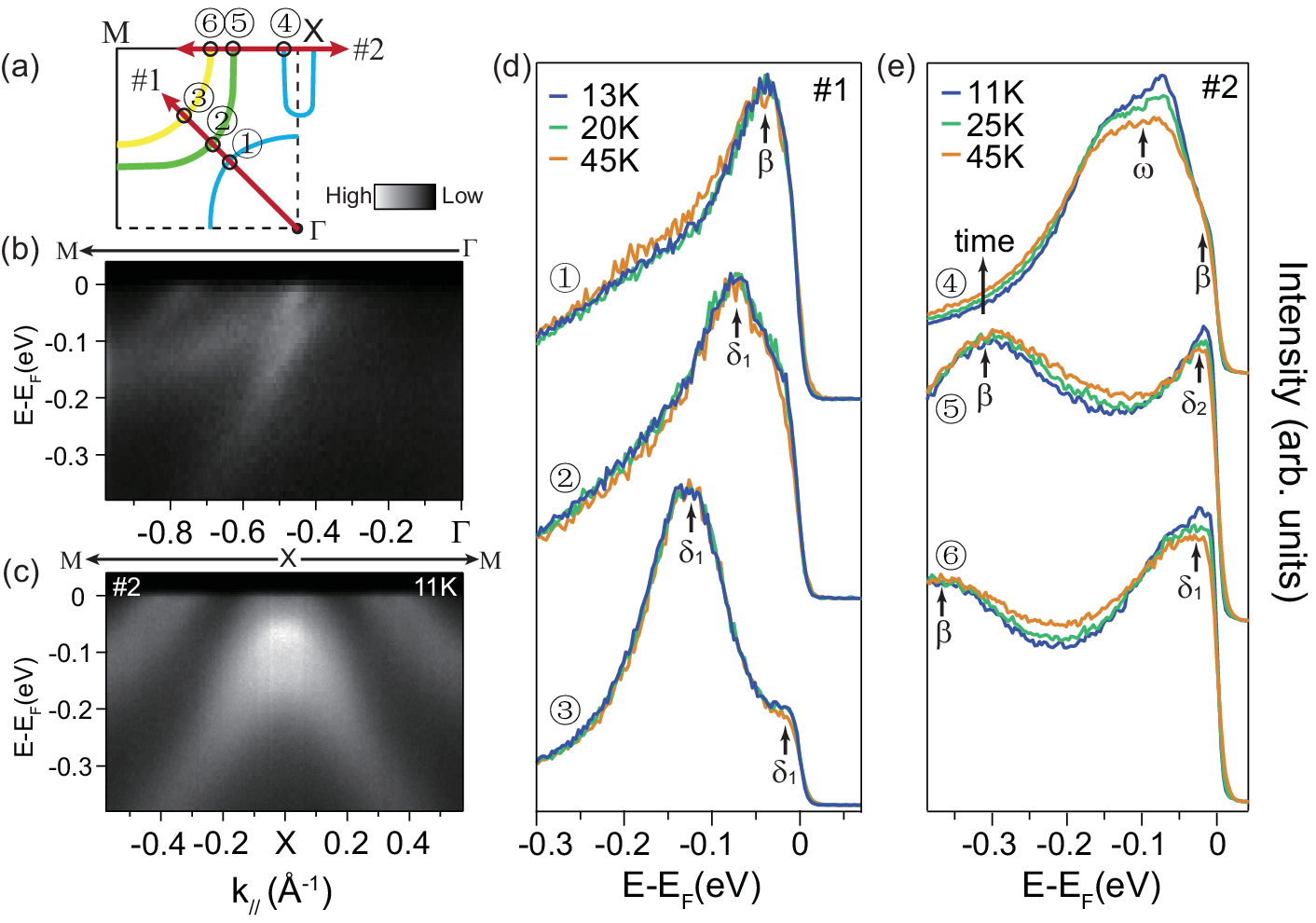}
\caption{(Color online)
(a) The sketch of a  Brillouin zone quardrant. (b) and (c)
Photoemission intensities along cuts \#1 and \#2 as indicated in panel
(a), respectively. (d) and (e) Temperature dependence of EDC¡¯s at various momenta
as marked in panel (a). The data in panels (b) and (d) were taken
with circularly polarized 75 eV photons at KEK, and the other data
were taken with randomly polarized 21.2 eV photons from a helium
discharge lamp.
}
\label{TemDep}
\end{figure}

One of the most intriguing aspects of KNi$_{2}$Se$_{2}$ is the possible heavy electronic state existing below $\sim$20 K ($T_H$), with an enhanced effective electronic band mass. However, this is not consistent with the well-defined bands observed at 15~K and the weakly correlated electronic structure.
To understand the origin of the large Sommerfield coefficient  in KNi$_{2}$Se$_{2}$, we compare its specific heat and electronic structure with those of BaNi$_{2}$As$_{2}$ and  KFe$_{2}$As$_{2}$.  The latter has a  Fe 3$d^{5.5}$ orbital configuration, which fosters stronger Hund's rule coupling and  higher electron correlations (renormalization factor $2\sim 4$) \cite{KFeAs ARPES}, leading to higher $\gamma$ (94.3 mJ mol$^{-1}$ K$^{-2}$) \cite{KFeAs HC}. Meanwhile, the $\gamma$  is 49.8 mJ mol$^{-1}$ K$^{-2}$ for  of KNi$_{2}$Se$_{2}$, and 13.2 mJ mol$^{-1}$ K$^{-2}$ for BaNi$_{2}$As$_{2}$ \cite{BaNiAs HC}.

As the electronic specific heat $\gamma$ coefficient is proportional to the DOS at $E_F$, which can be further estimated by the formula $\gamma\propto\sum\limits_n{}\int_{s_{n}(E_{F})}\frac{ds}{\mid  \triangledown _{k}E({k_{F})}\mid  }$, where $n$ is the band index, $\mid  \triangledown _{k}E({k_{F})}\mid$ is the mean velocity of an electron at $E_F$. For a 2D system, the formula can be simplified through: $d s$ $\to$ the perimeter of Fermi surface, and $\left | \triangledown _{k}E({k_{F})} \right |$ $\to$ the Fermi velocity or the band slope at $E_F$. Thus we can roughly evaluate the DOS at $E_F$ of these systems, as shown in Fig.~\ref{DOSestimation}.

By estimation, we notice that the total  Fermi surface perimeters are similar for all three systems  [Figs.~\ref{DOSestimation}(a), \ref{DOSestimation}(c), and \ref{DOSestimation}(e)]. However, the average Fermi velocity decreases rapidly in the order of  BaNi$_{2}$As$_{2}$, KNi$_{2}$Se$_{2}$ and KFe$_{2}$As$_{2}$ [Figs.~\ref{DOSestimation}(b), \ref{DOSestimation}(d), and \ref{DOSestimation}(f)].
Consequently, the estimated DOS value in KNi$_{2}$Se$_{2}$ is 4.0(6) times of that of BaNi$_{2}$As$_{2}$ , which is generally consistent with the corresponding ratio (3.8) of their $\gamma$ coefficients. Similarly, both the estimated DOS and $\gamma$ of   KFe$_{2}$As$_{2}$ are about 1.8$\sim$1.9 times of those of KNi$_{2}$Se$_{2}$. Therefore, our data suggest that the large $\gamma$ in KNi$_{2}$Se$_{2}$ is due to the relatively flat bands (especially $\delta_2$)  in combination with the large Fermi surface perimeters, instead of the heavy Fermion state. These are confirmed by our DFT  calculations, which gives flat $\delta_1$ and $\delta_2$. Our results also suggest that the large $\gamma$  in KFe$_{2}$As$_{2}$  is not due to heavy Fermion physics, but due to  strong correlations or Hund's rule coupling there as well. Furthermore, the remarkably good agreement between the estimated DOS at $E_F$ and the measured $\gamma$ for all three systems with different levels of correlation strength shows that this new way of quantitative analysis can provide valuable insight for understanding the thermal dynamical properties of these multi-band materials.

The neutron pair-distribution-function analysis revealed that the local CDW fluctuation occurs at $T_H$ and disappears upon further cooling \cite{KFeSe discovery}. Temperature evolution of the electronic structure is shown in Fig.~\ref{TemDep}. The photoemission intensity along $\mathit\Gamma$~-~\emph{M} (cut \#1) was displayed in Fig.~\ref{TemDep}(b), we found that there is no evident spectral weight change with temperature within our energy resolution, except for some thermal broadening effects  [Fig.~\ref{TemDep}(d)]. Along \emph{M}~-~\emph{X} (cut \#2), as shown in Fig.~\ref{TemDep}(c), there is a spectral weight suppression with increasing temperature for  $\delta_1$, $\delta_2$, and $\omega$, as indicated by the up arrows in Fig.~\ref{TemDep}(e). However, the spectral weight suppression rate do not alter noticeably across $T_H$. Similar spectral weight change was observed in Sr$_2$CuO$_2$Cl$_2$ and BaTi$_2$As$_2$O, which is explained by strong coupling between electrons and magnons or phonons \cite{SrCuOCl, BaTiAsO}.
For BaTi$_2$As$_2$O, a change of the spectral weight evolution rate was observed at the CDW ordering temperature. The absence of anomaly at $T_H$ here is likely due to the fact that the
CDW fluctuation in KNi$_{2}$Se$_{2}$ was reported to be dynamic and/or entirely uncorrelated between unit cells within the $ab$ plane, in contrast to the coherent CDWs observed in structurally related compounds such as NbSe$_2$ \cite{Dawei}. Nevertheless, since we observed spectral weight supression at nested Fermi surface sections, it may
indicate that the enhanced electron-phonon interactions in these sectors may be responsible for the local CDW fluctuations.

To summarize, we have systematically studied the electronic structure of KNi$_{2}$Se$_{2}$  by high-resolution ARPES. There are three bands intersecting the Fermi level, which form two electron-like pockets, one square-like and one elongated elliptical hole-like pockets surrounding the \emph{M}, $\mathit\Gamma$ and \emph{X} points, respectively. Furthermore,  the electronic structure of this multi-orbital superconductor shows relatively weak $k_z$ dependence, indicative of its 2D nature. Further comparison with the DFT band calculations suggests that the electron correlations therein are relatively weak. Our results clearly indicate that it is not the heavy Fermion behavior advocated before, but the relative small Fermi velocities of bands in combination with large Fermi surfaces that lead to the large DOS and electronic specific heat coefficient in  KNi$_{2}$Se$_{2}$.   We also practiced a new  way to quantitatively estimate DOS from ARPES data, which would  provide valuable insight for understanding the thermal dynamic properties.
Moreover, we  observed intriguing temperature dependence of the electronic structure in the nested Fermi surface sectors, although we did not observe evident anomaly   across   $T_H$. This may be consistent with the local fluctuating CDW in KNi$_{2}$Se$_{2}$.

\section{acknowledgement}

We gratefully acknowledge the helpful discussions with Prof. Xiangang Wan. This work was supported by National Basic Research Program of China (973 Program) under the grant Nos. 2011CBA00106, 2011CBA00112, 2012CB927401, and the National Science Foundation of China under Grant Nos. 11104304, 11227902 and 11274332. M. Y. Li and D. W. Shen are also supported by the $``$Strategic Priority Research Program (B)" of the Chinese Academy of Sciences (Grant No. XDB04040300).


\begin{references}
\bibitem{kz} We estimated the $k_z$ values according to the free-electron final-state model, where the inner potential of 15~eV was used.
\end{references}
\end{document}